%% file: main_dash22.tex
\pdfoutput=1

\documentclass[11pt]{article}

\usepackage{EMNLP2022}

\usepackage{times}
\usepackage{latexsym}

\usepackage[T1]{fontenc}

\usepackage[utf8]{inputenc}

\usepackage{microtype}

\usepackage{inconsolata}

\usepackage{graphicx}
\usepackage{enumitem}
\usepackage{multirow}
\usepackage{makecell}

\title{User or Labor: An Interaction Framework for Human-Machine Relationships in NLP}

\author{
    Ruyuan Wan \\ University of Notre Dame\\
    \texttt{rwan@nd.edu} \And
    Naome Etori \\  University of Minnesota \\
    \texttt{etori001@umn.edu} 
    \AND
  Karla Badillo-Urquiola \\ University of Notre Dame \\
  \texttt{kbadill3@nd.edu }  \And
  Dongyeop Kang \\ University of Minnesota \\ 
  \texttt{dongyeop@umn.edu}
    }

\begin{document}
\maketitle
\begin{abstract}

The bridging research between Human-Computer Interaction and Natural Language Processing is developing quickly these years. However, there is still a lack of formative guidelines to understand the human-machine interaction in the NLP loop. When researchers crossing the two fields talk about humans, they may imply a user or labor. Regarding a human as a user, the human is in control, and the machine is used as a tool to achieve the human's goals. Considering a human as a laborer, the machine is in control, and the human is used as a resource to achieve the machine's goals. Through a systematic literature review and thematic analysis, we present an interaction framework for understanding human-machine relationships in NLP. In the framework, we propose four types of human-machine interactions: Human-Teacher and Machine-Learner, Machine-Leading, Human-Leading, and Human-Machine Collaborators. Our analysis shows that the type of interaction is not fixed but can change across tasks as the relationship between the human and the machine develops. We also discuss the implications of this framework for the future of NLP and human-machine relationships.

\end{abstract}

\input{Sec1.Intro}
\input{Sec2.Related}

\input{Sec3.Method}

\input{Sec4.Properties} 
\input{Sec5.Relationships}

\input{Sec6.Discussion}
\input{Sec7.Limitations}
\input{Sec8.Conclusion}

\bibliography{anthology,custom}
\bibliographystyle{acl_natbib}

\appendix

\input{Sec9.Appendix}

\end{document}

%% file: Sec1.Intro.tex
\section{Introduction}
Research at the intersection of Natural Language Processing (NLP) and Human-Computer Interaction (HCI) is developing rapidly. Humans and machines are now both engaged in each step of the end-to-end NLP pipeline \cite{wang2021putting, wu2022survey}. NLP systems are trained on data created and annotated by humans, such as news articles \cite{da2019fine}, Wikipedia pages \cite{faruqui2018wikiatomicedits}, product reviews \cite{badlani2019disambiguating}, and social media posts \cite{joseph2021mis}. Yet, humans are also empowered by NLP systems, such as writing assistants \cite{chen2012flow}, collaborative text revision \cite{du2022read}, and translators \cite{gu2016learning}. Nonetheless, humans and machines play distinct roles in these scenarios. From the HCI perspective, researchers usually consider humans as the users of certain technology within the interaction, while many NLP researchers emphasize the labor responsibility of humans to improve the performance of NLP models in tasks. 


Humans and machines naturally have different strengths, such as trustworthiness, automation, and assessment ability \cite{shneiderman2020human, maes1995agents}. Humans stand out for trustworthiness, while machines are known for automation. Both humans and machines have the competence to evaluate each other, given their own specialties. 

While more and more interdisciplinary research and systems are being built to bridge the HCI and NLP fields, we still lack a normative understanding of how human and machine interaction works within the NLP context. More importantly, does the interaction work well? To fill this gap, we address two main research questions: 

\textbf{RQ1:} \textit{How does human-machine interaction happen in NLP?} 

\textbf{RQ2:} \textit{How do humans and machines interact with each other in NLP?} 

We address our research questions by conducting a systematic literature review on the interactive NLP research from leading HCI and NLP venues. Our goal was to define a generalizable human-machine interaction framework in NLP to explain current implementation, guide the design of human-machine interaction, and inspire future research in interactive NLP systems. Based on our synthesis, we defined three properties of interaction: \textit{continuity}, \textit{variety} of interaction options, and \textit{medium} (RQ1). We also conceptualized four relationships summarizing the roles of human and machine interaction patterns in different scenarios: \textit{Human-Teacher and Machine-Learner}, \textit{Machine-Leading}, \textit{Human-Leading}, and \textit{Human-Machine Collaborators} (RQ2). We used these properties and conceptualizations to contribute a theoretical interaction framework for human-machine relationships in NLP. 
Future research from HCI and NLP can use this framework to design and evaluate human-machine interaction for human and machine needs within their role and responsibility.

%% file: Sec2.Related.tex
\section{Related Work}
Traditionally, NLP models are trained, fine-tuned, and tested on existing datasets before being deployed to solve real-world problems.
While HCI research usually involves users designing, developing, implementing, and evaluating systems. To encourage collaboration between HCI and NLP, combining both approaches is required for successful 'HCI+NLP' applications \cite{HeuerHendrik2021MftD}. Further, \citet{HeuerHendrik2021MftD} proposed five methods for designing and evaluating HCI+NLP Systems: user-centered NLP, co-creating NLP, crowdsourcing, and user models.  

The human-machine interaction in NLP focuses on how people interact with machines and how NLP systems can be designed to support humans. In recent years, many researchers and practitioners have made significant advances in interactive NLP systems, such as text classification  \cite{godbole2004document}, text summarization \cite{passali2021towards}, semantic parsing and entity linking \cite{liang2020alice, he2016human, klie2020zero, zhong2020frustratingly}, dialogue systems \cite{sanguinetti2020annotating, liu2018dialogue}, topic modeling \cite{smith2018closing}.

\citet{wang2021putting} summarized recent human-in-the-loop NLP work based on their tasks, goals, human interactions, and feedback learning methods. According to \citet{wang2021putting}, a good human-in-the-loop NLP system must clearly communicate to humans what the model requires, provide user-friendly interfaces for collecting feedback, and effectively learn from it. For example, humans can provide various types of feedback, such as training data providers, annotations, and evaluators of the system's output to improve the model's performance, interpretability, or usability. 

Also, it is vital to understand the constraints of collecting and interpreting human inputs correctly. Many design considerations influence the efficiency and effectiveness of interactive learning from human feedback \cite{cui2021understanding}. For example, noise caused by human error (such as when a human teacher fails to provide the conventional ground truth) could be challenging in designing human-machine interaction systems. In addition, data collected from humans may be poor quality \citet{hsueh2009data}. Social bias \citet{fiske201913} can also be introduced automatically during data collection \citet{garrido2021survey} and language model training, which emerging the need of developing fair and responsible  models \citet{hutchinson2020social, mehrabi2021survey}.
Therefore, it is critical to appropriately interpret collected human data and analyze the effects of various interaction types on learning outcomes \cite{cui2021understanding}.

\input{Tables/table1.tex}

%% file: Tables/table1.tex
\begin{table*}[t]
\centering
\begin{tabular}{@{}c@{}c@{}}
\textbf{Relationships} &  \textbf{Papers}\\
\hline
\makecell{Human-Teacher,\\Machine-Learner}  & 
\makecell{\textbf{OUG} \citet{wiechmann2021activeanno}, \textbf{IUG} \citet{stiennon2020learning}, \\
\textbf{IVN} \citet{jandot2016interactive}, 
\textbf{IVM} \citet{wallace2019trick}, \\
\textbf{IVM}  \citet{liu2018dialogue}, \textbf{IVM} \citet{settles2011closing}, 
\textbf{IVM} \citet{godbole2004document}} \\
\hline
{Machine-Leading} & 
\makecell{\textbf{OUG} \citet{khashabi2020more}, 
\textbf{OUG} \citet{lawrence2018counterfactual}, \\
\textbf{IUG} \citet{lertvittayakumjorn2020find}, 
\textbf{IUG} \citet{he2016human}, \\
\textbf{IUN} \citet{liang2020alice}, 
\textbf{IVG} \citet{simard2014ice}, \\
\textbf{IVM} \citet{lo2020interactive}, 
\textbf{IVM }\citet{smith2018closing},
\textbf{IVM} \citet{ross2021evaluating}}\\
\hline

{Human-Leading}  &   
\makecell{\textbf{SUN} \citet{bhat2021people}, 
\textbf{SUN} \citet{rao2018learning}, \\
 \textbf{SVG} \citet{kim2021can}, 
 \textbf{SVM} \citet{coenen2021wordcraft}, \\
\textbf{IVM} \citet{chung2021talebrush}, 
\textbf{IVM} \citet{passali2021towards}} \\
\hline

\makecell{Human-Machine\\ Collaborators} &  
\makecell{\textbf{OUG} \citet{kreutzer2018can}, 
\textbf{OUN} \citet{khashabi2021genie}, 
\textbf{OVG} \citet{head2021augmenting}, \\
\textbf{SUN} \citet{ashktorab2021effects}, 
\textbf{IVG} \citet{karmakharm2019journalist}, \\
\textbf{IVN} \citet{hancock2019learning}, 
\textbf{IVN} \citet{van2021afriki}, 
\textbf{IVN} \citet{klie2020zero},\\
\textbf{IVM} \citet{clark2021choose},
\textbf{IVM} \citet{trivedi2019interactive}, 
\textbf{IVM} \citet{kim2019topicsifter},
}\\
\hline

\end{tabular}
\captionsetup{justification=centering}
\caption{Human-Machine Relationships Mapping Interaction Properties: The properties of the interaction in each paper are coded by the first letter of their three interaction properties: 
\textbf{O}/\textbf{S}/\textbf{I} represents \textbf{One-time}/\textbf{Sequential}/\textbf{Iterative}.
\textbf{U}/\textbf{V} represents \textbf{Unitary}/\textbf{Various}.
\textbf{G}/\textbf{N}/\textbf{M} represents \textbf{G}UI/\textbf{N}UI/\textbf{M}UI.
}
\label{table:papers}
\end{table*}

%% file: Sec3.Method.tex
\section{Survey and Analysis Methods}

We conducted a systematic literature review to understand the types of interactions humans and machines have within the NLP and HCI context. To select the targeted papers, we searched from ACL anthology (NLP database) and ACM Digital Library (HCI database) for articles that have been published over the last two years and included papers with keywords such as 'human-in-the-loop,' 'interactive,' 'collaborative,' 'active learning,' 'HCI + NLP,' 'human-machine,' and 'human-AI.' We also searched for workshops held in ACL or ACM conferences, such as HCI+NLP in EACL 2021\cite{blodgett2021proceedings}. Further, we conducted a backward reference search on the articles we found to identify any additional missing articles. This resulted in a total of 73 papers at the intersection of NLP and HCI. We narrowed down our search to only include articles that made algorithmic contributions, system contributions, or empirical contributions. We excluded papers that only contributed opinions, theories, surveys, or datasets. This resulted in a total of 54 papers. 

Next, we included only papers that had simultaneous interactions between humans and machines. We discarded 21 articles that had no interaction, or the interaction between the human and machine was asynchronous. This resulted in a final set of 33 articles included in our analysis. 
 
Further, We conducted a thematic analysis \cite{braun2012thematic} of our dataset. We began by reading through each article and taking notes on observations and insights regarding the interactions between humans and machines. We used these notes to develop a set of guiding questions that helped us generate our initial codes:
\begin{itemize}[noitemsep,topsep=1pt]
    \item \textit{What is the frequency of the interaction?}
    \item \textit{What are the different ways humans can interact with machines?}
    \item \textit{In what form does the interaction take place?}
    \item \textit{Who starts the interaction?}
    \item \textit{Who ends the interaction?}
    \item \textit{Who benefits from the interaction?}
\end{itemize}

%% file: Sec4.Properties.tex
\section{Findings}
After iterating on our codes informed by the guiding questions, we conceptualized the codes into two major dimensions: 1) the properties of human-machine interaction based on the first three questions, and 2) the types of human-machine relationships based on the last three questions. Table \ref{table:papers} summarizes the mapping between our dimensions, codes, and dataset. 

\subsection{Properties of Interaction }
Within our first dimension, we identified three major properties of how interactions happen, which address the first three guiding questions respectively: 1) continuity, 2) variety of interaction options, and 3) medium of interactions. 

\subsubsection{Continuity}
Continuity measures the frequency of interaction which can be one-time, sequential, or iterative, to perform a single task.
\paragraph{\textit{One-time interaction}} is when the human-machine interaction is designed to happen just once, usually in active learning. For example, ActiveAnno \cite{wiechmann2021activeanno} offers annotation generator functionality: human annotators will manually label documents at the beginning; if it reaches a threshold of the annotation generator, it will trigger the machine annotation generator to label the remaining documents in the project. 
\paragraph{\textit{Sequential interaction}} means that one agent acts first, and the other responds. Furthermore, the latter interactions are based on the previous exchanges. Over multiple rounds to complete the project, the interaction and the model will not update. 
For example, in the word guessing game to study the effects of communication directionality \cite{ashktorab2021effects}, the machine and human play as giver and guesser by giving word hints and word guessing together to let the guess win the game in 10 rounds. The human player is playing with the same machine player in each game, but previous rounds influence the words given in later rounds. 
\paragraph{\textit{Iterative interaction:}} the model performance improves through iterating over several interactions. For example, the Interactive Classification and Extraction (ICE) system enables humans to define the appropriate features through interactive features while humans can monitor their classifier's progress\cite{simard2014ice}. 
Researchers usually set up a limitation of the rounds of the interaction for experiments, but they show that theoretically, the continuous interaction (either sequential or iterative interactions) can continue with no end. 

\subsubsection{Variety of Interaction Actions }
Variety represents the number of ways humans can interact with machines. Some interactions are limited to a specific task (unitary), while some are flexible to multiple options (various). 
\paragraph{\textit{Unitary interaction}} is a single option of the interaction action, like labeling data \cite{wiechmann2021activeanno}, having humans perform one action to evaluate submissions \cite{khashabi2021genie}, ranking candidate questions \cite{rao2018learning}, giving natural language responses to query \cite{liang2020alice}, or selecting an answer from the multi-choice questions \cite{lertvittayakumjorn2020find, he2016human}. 

\paragraph{\textit{Various interaction}} is that there are multiple options for an interaction action. Humans can select what to do from multiple options, such as choosing continuation writing, rewriting, or filling in an interactive editor \cite{coenen2021wordcraft}. 
In CYOA~\cite{clark2021choose}, human has many options, such as deleting suggestions, submitting suggestions as-is, or writing a new suggestion.


\subsubsection{Medium of Interactions}
Medium in interactive NLP systems can be Graphical User Interfaces (GUI), Natural Language User Interfaces (NUI), or Mixed User Interfaces (MUI): 

\paragraph{\textit{Graphical User Interfaces (GUI)}} allow humans to select given options or highlight text. Such as, in FIND \cite{lertvittayakumjorn2020find}, humans answer multiple-choice questions about whether a given word cloud is relevant to a class to disable irrelevant hidden features.
Journalist-in-the-loop~\cite{karmakharm2019journalist} uses a web-based interface rumour analysis that takes user feedback. 
\cite{kreutzer2018can} collects reinforcement signals from humans using a 5-star rating interface. 

\paragraph{\textit{Natural Language User Interfaces (NUI)}} let the agent respond with natural language. For instance, GENIE~\cite{khashabi2021genie} uses text generation tasks. 
\citet{liang2020alice} 's ALICE utilizes contrastive natural language explanations to improve data efficiency in learning.

\begin{figure*}[!t]
\centering
\includegraphics[width=0.88\textwidth,trim={0.5cm 0.5cm 0.5cm 0.5cm},clip]{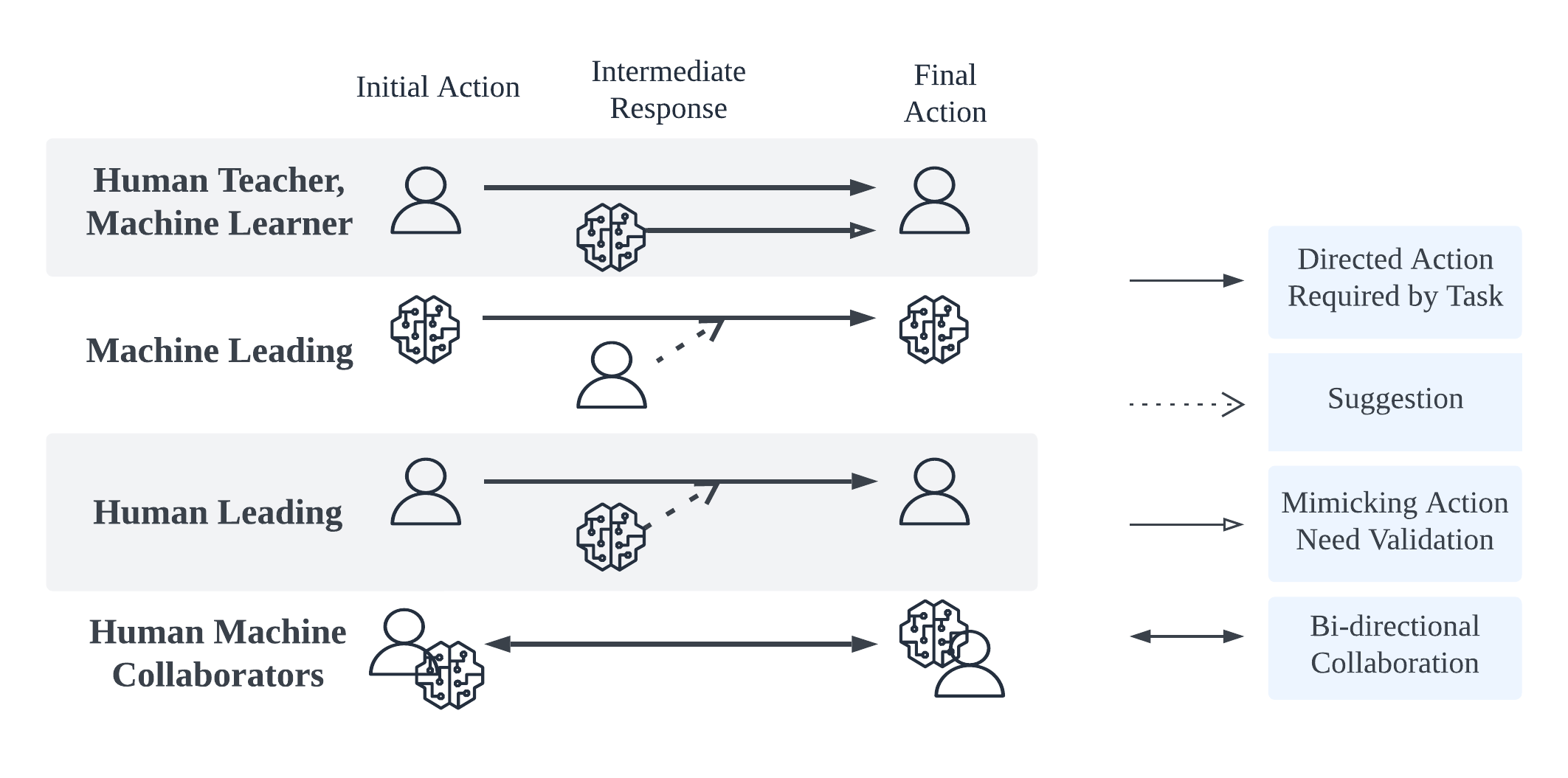}
\caption{Interaction Framework for Human-Machine relationships in NLP
\label{fig:human_machine_relation_type}
}
\end{figure*}

\paragraph{\textit{Mixed User Interfaces (MUI)}} contain both graphical and natural language interfaces. For example, in the Interactive NLP in Clinical Care \cite{trivedi2019interactive}, physicians can add or remove highlighted predicted sentences in a report and write natural language feedback in a separate box. 
Also, Wordcraft allows humans to start writing with a prompt and change text and selection options for machine editing in the side GUI \cite{coenen2021wordcraft}.
In CYOA~\cite{clark2021choose}, a human can write a storyline alone and delete suggestions, while models provide suggestions of the story, and then the human can submit the suggestion. 
Participants are asked to score on a Likert-scale and open-ended questions about the systems and suggestions they received after submitting their stories.
In DUALIST~\cite{settles2011closing} humans can label documents by clicking the appropriate class from the drop-down menus below each text, 
In addition, each column has a text box where users can "inject" domain knowledge by typing in random words. 
Users must click a large submit button at the top of the screen to retrain the classifier and get a new set of queries. 
Smoothed dictionary features~\cite{jandot2016interactive} use a methodology to solicit features from humans or teachers. 
In ~\citet{passali2021towards} human has the option to view visualization and choose different decoding strategies. 

%% file: Sec5.Relationships.tex
\subsection{Relationships of Human and Machine}
For our second dimension, we conceptualized four relationships (each based on the last three guiding questions): 1) \textit{Human-Teacher and Machine-Learner}, 2) \textit{Machine-Leading}, 3) \textit{Human-Leading}, and 4) \textit{Human-Machine Collaborators}.

\paragraph{\textit{Human-Teacher and Machine-Learner}}
When humans initiate the interaction, then machines learn from humans to mimic the task, after that, humans evaluate and give feedback on the machines' learning results in the end. Through this interaction, humans benefit from machines' automation in finishing tasks, and machines also benefit from learning to improve their performance.   

For instance, humans annotate and label data at the beginning in ActiveAnno (\citet{wiechmann2021activeanno}), ICE (\citet{simard2014ice}), DUALIST~ \cite{settles2011closing} and HIClass~\cite{godbole2004document}. Once the machine has learned enough from the human annotation, it can predict labels for the remaining documents.
In Trick-Me-If-You-Can~\cite{wallace2019trick}, human authors are guided to break the model by writing adversarial questions, and the machine exposes the predictions and interpretations of the answers to humans. 
In smoothed dictionary features~\cite{jandot2016interactive}, the machine elicits dictionary features from a teacher or human. 

\paragraph{\textit{Machine-Leading}}
Machines initiate interactions with their optimal competence, then humans respond with suggestions, and later, machines iterate on the task based on the intermediate suggestions. From the interaction, machines benefit from humans' knowledge in improving their own capacity, while humans don't earn anything from the process.

For instance, ALICE \cite{coenen2021wordcraft}, FIND \cite{lertvittayakumjorn2020find}, Interactive NLP in clinical care \cite{trivedi2019interactive}, and Human-in-The-Loop Parsing \cite{he2016human}, have a baseline model to perform the corresponding NLP tasks. 
Then human steps into interactively selecting useful features, promoting machine learning efficiency. 
In Interactive Entity Linking~\cite{lo2020interactive}, an instance chooses mentions for human annotation using an active learning approach. The goal is to improve entity linking accuracy while updating the embedding model. The machine triggers the interaction, the human assists the machine in the annotation process, and human annotation feedback is used to update the model. Finally, high machine accuracy determines the end of the interaction process. 

\paragraph{\textit{Human-Leading}}
Humans initiate the task, then machines give suggestions based on their pre-trained expertise, later, humans select the way they want to interact with machines. Via the  interaction, humans gain help from machines, but machines don't take any benefit from these interactions.    

Like Wordcraft (\citet{coenen2021wordcraft}), human plays the lead role because human triggers the interaction by writing a prompt, selecting collaborative writing options, and eventually deciding what to write and when to end. 
In Towards Human-Centered Summarization~\cite{passali2021towards}, human actively participates and takes a leading role in the summarization process, such as deciding on the decoding strategies during the inference stage, choosing visualization color and viewing visualization of the attention weights, combining sentences from the various summaries to create a new one that can be used in fine-tuning the model.  

\begin{figure*}[!t]
\centering
\includegraphics[width=0.94\textwidth,trim={0.3cm 0.57cm 0.3cm 0.5cm},clip]{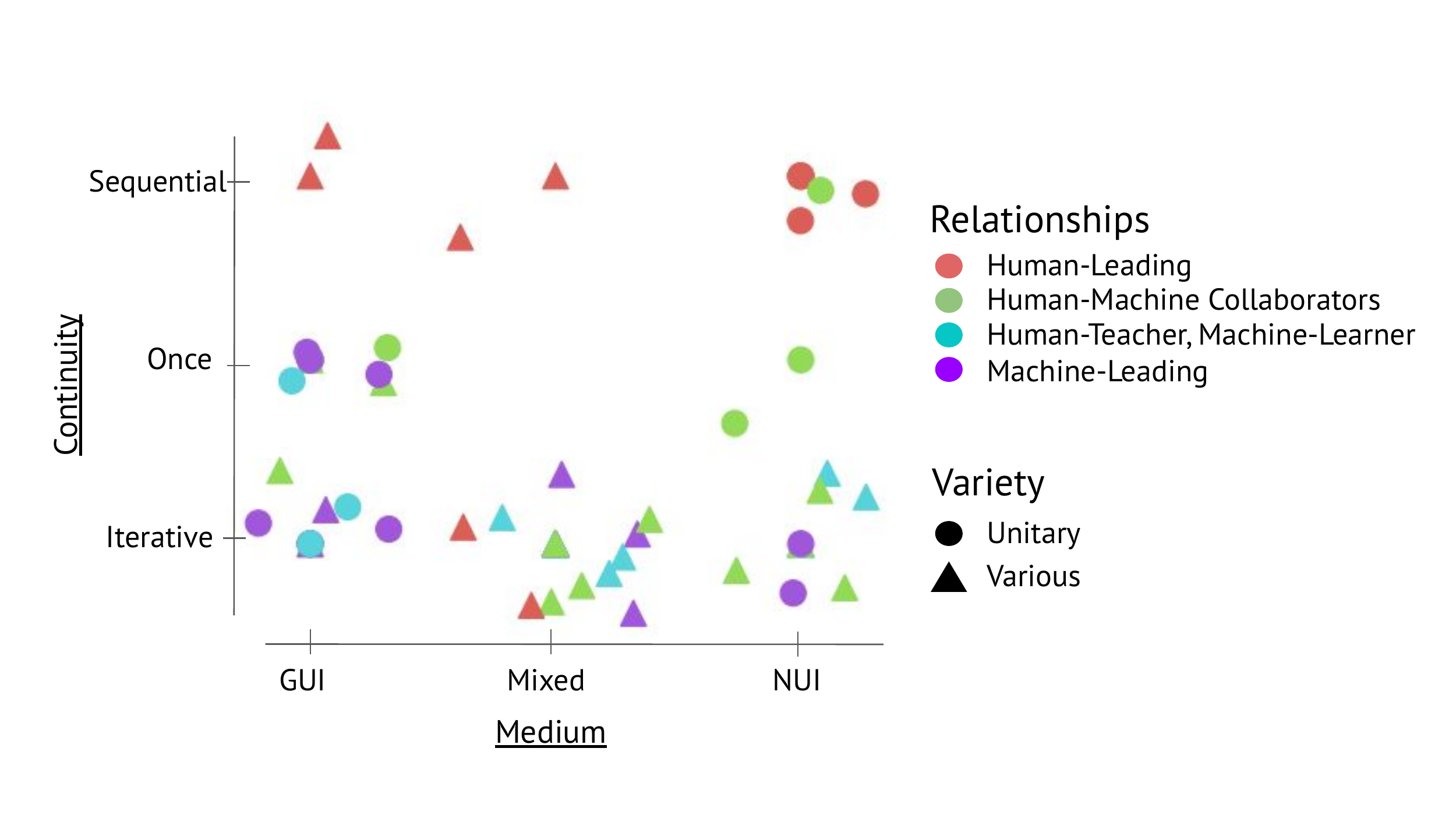}
\caption{Human-Machine Relationships Mapping Interaction Properties
\label{fig:relationship_mapping}
}
\end{figure*}

\paragraph{\textit{Human-Machine Collaborators}}
Either a human or machine initiates the task, then the other one gives the response. There is no explicit benefit for humans or machines during the interaction. 

For example, in \citet{ashktorab2021effects}, humans and machines can play as giver and guesser and cooperate for a common goal: to help the guesser answer the right word. In GENIE~\cite{khashabi2021genie}, human annotators work with a machine to assess leaderboard submissions on various axes such as correctness, conciseness, and fluency and compare their responses to a variety of automated metrics. Journalist-in-the-loop~\cite{karmakharm2019journalist} is a rumor annotation service that continuously allows journalists to provide feedback on social media posts. The feedback is then used to improve the neural network-based rumor classification mode.
Also, human-machine collaboration improves the journals' productivity. 
In AfriKI~\cite{van2021afriki}, human authors collaborate with machines to generate Afrikaans poetry through phrase selection and vertically arranging poetry lines. The collaboration promotes human creativity and also improves the robustness of the model. From Zero to Hero~\cite{klie2020zero}, an interactive machine learning annotation supports that guides the user in locating text entities and deciding on the correct knowledge base entries. As a result, the annotation speed and quality improve tremendously and reduce human cognitive load. 
In CYOA~\cite{clark2021choose}, humans and machines take turns writing a story, and two models generate suggestions for the following line after the human writes the story's first line. The human suggestions provide interpretability of the model performance and improve human creativity.

%% file: Sec6.Discussion.tex
\section{Discussion}
Based on our findings, we developed an interaction framework for the four types of human-machine relationships in one interaction cycle (see Figure \ref{fig:human_machine_relation_type}). The "human-in-the-loop" concept has become popular in NLP. But it is too broad to describe the nuance in similar but different human-machine interactions. Our interaction framework depicts human and machine actions under different roles.

It is important for NLP and HCI practitioners to define human-machine interactions when designing interactive NLP systems because it builds up a clearer understanding of the human and machine's strengths and weaknesses. We can use the identified strengths to complement the weaknesses. For example, in \textit{Human-Teacher and Machine-Learner interactions}, the strengths of automation in annotation tasks can be leveraged to complete a task more quickly, while the human's strength in assessment can validate the machine's output. For \textit{Human-Leading} or \textit{Machine-Leading}, the leader may have limited knowledge and capacity that could be complemented with external expertise. 

Next, we share how NLP and HCI practitioners can use our conceptual framework in practice. While \textit{Human-Leading} is similar to \textit{Human-Teacher, Machine Learner} (since both are driven by humans who take initial and final action with machines' intermediate response), the framework captures the different statuses of machines and the corresponding actions of humans and machines. As a learner, the machine is a novice at the beginning and is launched by mimicking humans finishing their remaining tasks. However, the machines in the \textit{Human-Leading} relationship have their own expertise to offer suggestions. Accordingly, this also leads to different humans' final actions, i.e., humans as teachers will validate machines' learning results, while humans would choose machines' suggestions based on personal interests during human-leading interaction. 

Additionally, the framework can be used as a guideline to check whether the interaction design is appropriate. For instance, the machine directly overwrites a human's control will be problematic when \textit{Human-Leading} is required. For instance, someone is writing in colloquial English, using 'ain't' as a general preverbal negator \cite{rickford1999african}. A machine trained in standard English will detect this as an error and would like to correct it, but it violates the writer's intention.

Figure \ref{fig:relationship_mapping} shows some trends of interactions: \textit{Human-Leading} relationship usually happens with sequential interactions;  \textit{Machine-Leading} relationship uses more GUI and has iterative interaction; \textit{Human-Machine Collaborators} mainly use NUI; \textit{Human-Teacher, Machine-Learner} relationship usually happens iteratively as well.


%% file: Sec7.Limitations.tex
\section{Limitations}
From the papers we reviewed, we didn't find any interaction that machines validate humans' actions. This might be because humans overall are more trustworthy than machines and may be influenced by the range of literature we reviewed. Additionally, 'collaboration' can sometimes be interchangeable with 'interaction' based on their semantic meanings. But we strive to name our precise definitions of each relationship with clear and straightforward enough abstract names.  

In addition, we initially coded \textit{'what are the outcomes of the research paper'} to understand how interaction can influence the research outcome, such as efficiency, creativity, etc. But it didn't synthesize sufficiently with other codes. More importantly, the research outcome is usually the goal of those research papers so we couldn't derive the causal relationship from interaction design to the research outcome. However, this guides us in the future to study how we can manipulate desired research improvement by designing human-machine interaction. 

%% file: Sec8.Conclusion.tex
\section{Conclusion}
Humans and machines interact with each other in a variety of ways. For example, humans may be involved in providing input to a machine learning algorithm, labeling data for training purposes, or evaluating the output of a machine learning system. Additionally, humans may interact with machine learning systems through natural language interfaces, such as chatbots or virtual assistants. We contribute an interaction framework for human-machine interactions through a systematic literature review and thematic analysis, which conceptualizes four human-machine relationships based on three different interaction properties, to help researchers and practitioners better understand and manage human-machine interactions in NLP.
